\documentclass[amsmath,amssymb,twocolumn,showpacs,prb]{revtex4}
\usepackage[dvips]{graphicx}
\usepackage{bm}

\begin{document}
\title{
Fluctuation theorem for a double quantum dot coupled to a point-contact electrometer}

\author{
D. S. Golubev$^1$,
Y. Utsumi$^2$,
M. Marthaler$^{3}$,
and
Gerd Sch\"on$^{1,3,4}$
}

\affiliation{
$^1$ Institut f\"ur Nanotechnologie, Karlsruhe Institute of Technology, 76021 Karlsruhe, Germany \\
$^2$ Department of Physics Engineering, Faculty of Engineering, Mie University, Tsu, Mi-e, 514-8507, Japan
\\
$^3$ Institut f\"{u}r Theoretische Festk\"{o}rperphysik, Karlsruhe Institute of Technology, 76128 Karlsruhe, Germany\\
$^4$ DFG Center for Functional Nanostructures (CFN), Karlsruhe Institute of Technology, 76128 Karlsruhe, Germany
}

\begin{abstract}
We study single-electron transport through a double quantum dot (DQD) monitored by a capacitively coupled quantum point-contact (QPC) electrometer.
We derive the full counting statistics for the coupled DQD - QPC system and
obtain 
the joint probability distribution of the charges transferred through the DQD and the QPC consistent with the fluctuation theorem (FT).
The system can be described by a master equation with tunneling rates depending of the
counting fields and satisfying a generalized local detailed-balance relation.
Furthermore, we derive universal relations between the non-linear corrections to the current and noise,
which can be verified in experiment.
\end{abstract}

\date{\today}
\maketitle

\newcommand{\mat}[1]{\mbox{\boldmath$#1$}}
\newcommand{\mtau}{\mbox{\boldmath$\tau$}}
\newcommand{\cgf}{{\cal W}}

\section{Introduction}

The well known fluctuation-dissipation relation~\cite{Kubo} between the conductance
of a mesoscopic system and its noise is fundamentally important but limited to the low-bias, linear response regime.
The fluctuation theorem (FT)~\cite{Evans,Lebowitz,Andrieux1,Tobiska,Esposito,Atsumian,AndrieuxA,Foerster,SaitoU,Andrieux2,Esposito1,Altland1,Campisi,DSanchez,RSanchez,BulnesCuetara,Hayakawa}
extends this relation to the high-bias and non-linear regime.
It is formulated in terms of the entropy production in a mesoscopic system $\Delta S$ during time $\tau$ 
and relates the probabilities $P_\tau(\mp\Delta S)$ to find negative or positive
values, $\mp\Delta S$, in the simple relation  
\begin{eqnarray}
{P_\tau(-\Delta S)}/{P_\tau(+\Delta S)}= e^{-\Delta S}.
\label{FTS}
\end{eqnarray}
The FT (\ref{FTS}) has been verified in chemical, biophysical~\cite{Wang,Carberry}, and condensed matter~\cite{Garnier,Schuler} experiments. 
Recently it has also been tested in low-temperature electron transport experiments 
with a serially coupled double quantum dot (DQD)~\cite{Utsumi1}, and with an Aharonov-Bohm interferometer~\cite{Nakamura}. 
In these cases the entropy production is related to the Joule heat $\Delta S =q eV_S/ T$, where $q$ is the number of electrons transmitted through the DQD during the measurement time $\tau$, $V_S$ is the source-drain voltage bias.
Hence the FT (\ref{FTS}) takes the form 
\begin{eqnarray}
\frac{P_\tau(-q)}{P_\tau(q)}=\exp
\left( \! 
- \frac{qeV_S}{ T}
\! \right).
\label{FT}
\end{eqnarray}

Experimental tests of the FT (\ref{FT}) in mesoscopic conductors are challenging 
because of the  usually  strong coupling to the environment, 
and, on top of that, the back action of the charge detector, which is relatively enhanced at low temperatures.  
In this paper we derive the full counting statistics (FCS) of the charge transport through a DQD
capacitively coupled to a quantum  point contact (QPC), which serves as a detector. 
We study the effect of both the electromagnetic environment and the back action of the QPC electrometer on the transport properties of the DQD.  
In earlier work \cite{Utsumi1} we had shown that the QPC back action leads to a violation of the FT (\ref{FT}) 
when only the two-terminal DQD is considered. 
Here we explicitly show that the combined system of the DQD and the QPC satisfy a four terminal version of the FT  
\begin{eqnarray}
\frac{
P_\tau(-q,-q')
}{
P_\tau(q,q')
}
=
\, 
\exp
\left(
- 
\frac{ q eV_S + q' eV_{\rm QPC} }{ T}
\right). 
\label{fullft} 
\end{eqnarray}
Here $P(q,q')$ is the joint probability distribution for $q$ charges to be transferred through the DQD 
and $q'$ charges through the QPC during the time $\tau$, and $V_{\rm QPC}$ is the QPC bias voltage.
We reveal that the FT (\ref{fullft}) is the consequence of generalized local detailed balance relations, Eqs.~(\ref{eldb1}-\ref{eldb3}) presented below, 
between the DQD transition rates, which turn out to depend on the counting field of the QPC. 

The direct experimental test of the relation (\ref{fullft}) is difficult
because the rate of electron tunneling through the QPC is too fast and
the measurement of the joint probability distribution $P_\tau(q,q')$ is impossible.
Therefore, similarly to Ref.~\onlinecite{RSanchez},
we propose to test the relations between the non-linear transport coefficients, Eqs.~(\ref{universal1}) and (\ref{universal2}).
These relations are the consequence of the FT (\ref{fullft}) and can be verified by measuring the currents and the noises of the DQD and the QPC  
as functions of the bias voltages $V_S$ and $V_{\rm DQD}$.

The FT (\ref{fullft}) has been previously discussed in Refs.~\onlinecite{SaitoU,Atsumian,AndrieuxA} at a formal level. 
It has been studied in detail for a system of two capacitively coupled quantum dots~\cite{RSanchez,BulnesCuetara}, 
both of which are fully described by master equations. 
Here we explicitly derive it for the specific case of coupled DQD and QPC. 
Our analysis requires involved calculations an is based on the separation between the time-scales relevant for the dynamics of the DQD,
which is a slow classical system described by a master equation, and of the QPC, which is a fast quantum system, whose dynamics we describe in a path integral approach. 
As a result, the combined system is described by a master equation with tunneling rates modified by the QPC back action.
We take into account two back action mechanisms: 
(i) back action due to the switching of the QPC current between several values corresponding to different DQD charging states, and 
(ii) back action due to the electromagnetic environment excited by QPC shot noise.   
An approach similar to ours, which, however, ignores (ii), has been earlier developed by Braggio {\it et al.}~\cite{Braggio2}

The paper is organized as  follows: 
In Secs.~\ref{sec:model1} 
we present the experimental setup and  our main results. 
In Sec.~\ref{sec:model} we specify the full microscopic model, then derive in 
Sec.~\ref{sec:cf} the master equation for the DQD, and use it
to find the FCS of the combined DQD + QPC system. 
Finally, in section~\ref{sec:summary} we summarize our results.

\section{Model and Main Results}
\label{sec:model1}
\subsection{Setup}

\begin{figure}[ht]
\includegraphics[width=.95 \columnwidth]{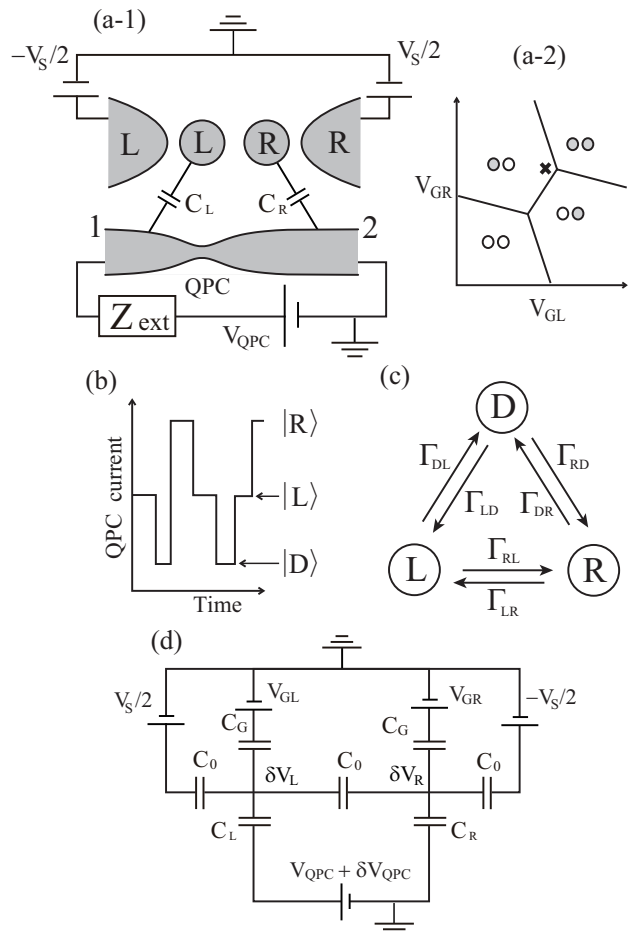}
\caption{
(a-1) Setup used to count single-electron tunneling events through the DQD in real time.
The impedance $Z_{\rm ext}(\omega)$ models the electromagnetic environment which affects the QPC transport properties.   
(a-2) Low bias stability diagram of the double quantum dot. The solid lines indicate the equality of the energies
of the neighboring charging states. The operating point
is indicated by a black cross. It is chosen in the vicinity of a tricritical point, 
where the energies of the three charging states $| L \rangle$, $| R \rangle$, and $| D \rangle$ are equal to each other. 
(b)
The QPC current switches between three values corresponding to the three charge states of the DQD. 
(c) 
The relevant transition processes. 
Circles represent the double-dot states and arrows represent the directions of the transitions. 
(d) 
An equivalent circuit of the system. 
}
\label{fig:setup}
\end{figure}

We consider a DQD capacitively coupled to a QPC, which serves as a charge detector~\cite{Fujisawa,Ihn}. 
The schematics of the system, as used in the experiment by Fujisawa {\it et al.}~\cite{Fujisawa}, is shown in Fig.~\ref{fig:setup} (a-1). 
Electrons tunnel between the two quantum dots and the leads via high-resistance tunnel junctions.
The level spacing in the quantum dots is large enough such that only one level in each dot participates in the charge transport. 
The gate voltages applied to the dots are tuned in such a way that only three charge states of the DQD contribute to the transport: 
the state with occupied left and empty right dots (state $|L\rangle$), the state with
empty left and occupied right dot (state $|R\rangle$) and, finally, the state with both dots occupied (state $|D\rangle$), see Fig.~\ref{fig:setup} (a-2). 
We note that our analysis equally applies to the case when the states $|L\rangle$, $|R\rangle$,
and $|0\rangle$ (both dots empty) are occupied, i.e. in the vicinity of another tricritical point in Fig..~\ref{fig:setup} (a-2).  
The state of the DQD is characterized by the vector ${\bf p}$ of the occupation probabilities
of these states,
\begin{eqnarray}
{\bf p}^T
=
\left( p_L,p_R,p_D\right).
\label{p} 
\end{eqnarray} 
We assume spin degeneracy and from now on ignore the spin degrees of freedom. 
It is justified because the tunnel coupling between the leads and the dots is supposed to be small, 
and hence the non-trivial spin related physics, such as Kondo effect, 
is irrelevant.

Due to the coupling to the DQD, the QPC transmission and, as a consequence,
the QPC current switches between three values corresponding to the DQD charge states $|R\rangle,|L\rangle$ and $|D\rangle$, 
see Fig.~\ref{fig:setup} (b).
We denote the corresponding QPC transmissions as ${\cal T}_{|R\rangle},{\cal T}_{|L\rangle},{\cal T}_{|D\rangle}$
and assume the following relation between them
${\cal T}_{|\! D \rangle} < {\cal T}_{|\! L \rangle} < {\cal T}_{|\! R \rangle} $.
In addition, we assume that the QPC is tuned to the tunneling regime, where all transmission are small, ${\cal T}_{|n\rangle}\ll 1$.

Previous experiments revealed that  the intrinsic capacitive coupling between the QPC detector and the DQD leads to a strong parasitic back action of the former on the DQD transport \cite{Aguado,Utsumi1}. 
The mechanism of this backaction is as follows: 
The fixed, i.e. time-independent, QPC bias voltage $V_{\rm QPC}$ is applied to and the QPC 
and the serially connected environment impedance $Z_{\rm ext}(\omega)$ 
(see Fig. \ref{fig:setup} (a-1)). 
Since both of these elements produce noise, the instantaneous voltage drops 
across the QPC, $v_{\rm QPC}(t)$,
and across the external impedance, $v_{\rm ext}(t)$, fluctuate in time, 
while their sum stays constant, $v_{\rm QPC}(t)+v_{\rm ext}(t)=V_{\rm QPC}$.
Through the capacitive coupling, the fluctuations of the voltage across the QPC are transferred
to the DQD and act in the same way as fluctuations of the gate potentials applied to each of the two quantum dots.
These fluctuations, in turn, modify the tunneling rates \cite{Aguado}. 

The setup shown in Fig.~\ref{fig:setup}(a-1) allows one to distinguish the direction of the electron tunneling,
and, hence, is suitable for testing the FT (\ref{FT}) provided the back action from the QPC charge
detector is minimized. 
For example, the switching of the QPC current between the levels  $|L\rangle$ and $| R \rangle$ 
corresponds to the transfer of one electron from the left  to the right dot, 
while the switching $|R \rangle \to |L \rangle$ signals a transfer in the opposite direction. 
Below we will show that the setup of  Fig.~\ref{fig:setup}(a-1) also allows
testing the extended version of the FT (\ref{fullft}).

\subsection{Full counting statistics and generalized master equation}

In this subsection  we summarize our main findings, while the details of the derivation can be found in Sec.~\ref{sec:mastereq}. 
The coupled system of the DQD and the QPC is described by
the joint probability  distribution $P_\tau(q,q')$ of the charges $q$ and $q'$ being 
transfered through the DQD and the QPC, respectively. 
We introduce the corresponding characteristic function (CF) 
\begin{eqnarray}
{\cal Z}_\tau(\lambda,\chi)
=\sum_{q,q'}
P_\tau
(q,q') \, 
{\rm e}^
{
i \lambda q
+
i \chi q'
}, 
\label{eqn:cf}
\end{eqnarray}
and the cumulant generating function (CGF) 
\begin{eqnarray}
{\cal F}(\lambda,\chi)
=
\lim_{\tau \to \infty}
\frac{1}{\tau}
\ln 
{\cal Z}_\tau(\lambda,\chi)
, 
\label{eqn:cgf}
\end{eqnarray}
where $\lambda$ and $\chi$ are the counting fields of the DQD and QPC respectively. 

The charge transport though the DQD occurs by sequential tunneling of electrons through the high-resistance tunnel junctions. 
This process is described by a simple master equation for the occupation probabilities of the three DQD charge states. 
Since we are interested in the full counting statistics of the electron transport, we need to  keep track of the counting fields $\lambda$ and $\chi$.
In this case the occupation probabilities (\ref{p}) depend on the counting fields (see Eq.~(\ref{p_}) for precise definition),
and the master equation acquires the form
\begin{eqnarray}
\partial_\tau
{\bf p}(\tau)
=
\mat{M}(\lambda,\chi) 
\, 
{\bf p}(\tau), 
\label{master}
\end{eqnarray}
where $\mat{M}$ is a $3 \times 3$ transition matrix, 
\begin{widetext}
\begin{eqnarray}
\mat{M}
\! &=& \!
\left[
\begin{array}{ccc}
-{\Gamma}_{RL}(0)-{\Gamma}_{DL}(0)  & \Gamma_{LR}(\chi) \, {\rm e}^{-i\lambda} & \Gamma_{LD}(\chi) \\
\Gamma_{RL}(\chi) \, {\rm e}^{i\lambda} & -{\Gamma}_{LR}(0)-{\Gamma}_{DR}(0) & \Gamma_{RD}(\chi) \\
\Gamma_{DL}(\chi) & \Gamma_{DR}(\chi) & -{\Gamma}_{LD}(0)-{\Gamma}_{RD}(0) 
\end{array}
\right]
+\,\left[
\begin{array}{ccc}
{\cal F}_L+\delta{\cal F}_0 & 0 & 0 \\
0 & {\cal F}_R+\delta{\cal F}_0 & 0 \\
0 & 0 & {\cal F}_D+\delta{\cal F}_0
\end{array}
\right]. 
\label{M}
\end{eqnarray}
\end{widetext}
From the solution of the rate equation (\ref{master}) the CF (\ref{eqn:cf}) is found as follows 
\begin{eqnarray}
{\cal Z}_\tau (\lambda,\chi)
=
\sum_{n=L,R,D}
p_n(\lambda,\chi;\tau).
\label{eqn:cfp}
\end{eqnarray}
The outlined procedure differs from the usual one~\cite{Bagrets} only 
by the dependence of the tunnel rates $\Gamma_{mn}$ on the QPC counting filed $\chi$. 

Let us now discuss various parameters which appear in the transition matrix (\ref{M}).
The parameters ${\cal F}_n$ are the CGF of the QPC  at a fixed charging state of the DQD, indicated by the subscript $n$. 
They are given by 
\begin{eqnarray}
{\cal F}_n(\chi)=
\frac{ {\cal T}_{|n \rangle} }{\pi}
eV_{\rm QPC}
\left[ \frac{e^{-i\chi}-1}{e^{eV_{\rm QPC}/T}-1} + 
\frac{e^{i\chi}-1}{1-e^{-eV_{\rm QPC}/T}}  \right].
\nonumber \\
\label{cgfqpc}
\end{eqnarray}
The correction $\delta {\cal F}_0(\chi)$ is defined in Eq. (\ref{cgfgauss}) 
and is due to influence of the environment 
on the QPC.

The transition rates in the matrix (\ref{M}) are defined as follows  
\begin{eqnarray}
\Gamma_{LR}(\chi) &=& 2 \pi  t_c^2 P_{dd,\chi}(\Delta_R-\Delta_L) \, ,
\label{glr}
\\
\Gamma_{RL}(\chi) &=& 2 \pi  t_c^2 P_{dd,\chi}(\Delta_L-\Delta_R) \, ,
\label{grl}
\\
\Gamma_{DR}(\chi) &=& \Gamma_L \int d E  f(-\Delta_R-\mu_L+E) P_{L,\chi}(E),
\label{gdr}
\\
\Gamma_{RD}(\chi) &=& \Gamma_L \int  d E  [1-f(-\Delta_R-\mu_L-E)] P_{L,\chi}(E),
\nonumber \\
\label{grd}
\\
\Gamma_{DL}(\chi) &=& \Gamma_R \int  d E f(-\Delta_L-\mu_R+E) P_{R,\chi}(E),
\label{gdl}
\end{eqnarray}
\begin{eqnarray}
\Gamma_{LD}(\chi) &=& \Gamma_R \int  d E [1-f(-\Delta_L-\mu_R-E)] P_{R,\chi}(E).
\nonumber \\
\label{gld}
\end{eqnarray}
Here we introduced the hopping matrix elements between the two quantum dots $t_c$; 
the energy differences $\Delta_n=E_n-E_D$ between the energies $E_n$ ($n=L,R,D$) of the DQD charging states; 
the tunnel coupling strength between the dots and the leads $\Gamma_L,\Gamma_R$;  and the Fermi function 
$f(E)=1/({\rm e}^{E/T}+1)$. 
The back action of the QPC on the transport properties of the DQD is encoded
in the functions $P_{n,\chi}(E)$, where $n=dd,L,R$ indicates one of the three tunnel barriers
in the DQD. 
These functions describe both the effect of the QPC shot noise and the excitation of the QPC environment~\cite{Ingold,Aguado} 
and depend on the QPC counting field $\chi$. 
They are defined as
\begin{eqnarray}
P_{n,\chi}(E)
\! &=& \!
\int \!\! \frac{dt}{2 \pi} \, 
{\rm e}^{i \kappa_n^2 \Phi_\chi(t)+iE \, t} \, . 
\label{pe}
\end{eqnarray}
The dimensionless parameters $\kappa_L=-\kappa_R=\kappa_{dd}/2$ 
characterize the capacitive coupling between the DQD and the QPC and are found from the equivalent circuit shown in Fig. ~\ref{fig:setup} (d):
\begin{eqnarray}
\kappa_{dd}=
-\frac{C_L (C_0+C_G+C_R)}
{ 3 C_0^2 + (C_G+C_L) (C_G+C_R) + 2 C_0 \tilde C },
\end{eqnarray}
where $\tilde C=2 C_G+C_L+C_R$. The function $\Phi_\chi(t)$ reads
\begin{eqnarray}
\Phi_\chi(t)  &=& i e^2\int  d \omega |Z_t(\omega)|^2
\nonumber\\ &&\times \,
\frac{e^2{\cal F}_0(\chi)+S_{0}(\omega)-{\rm e}^{-i \omega t}S_{\chi}(\omega)}{\omega^2 \Omega_\chi(\omega)},
\label{Phi1}
\end{eqnarray}
where
\begin{eqnarray}
&& \Omega_\chi(\omega) = 1
-
\frac{|Z_t(\omega)|^2}{\omega^2}
\big[
S_{\chi}(\omega) S_{\chi}(-\omega)
\nonumber\\ &&
-\, \big(S_{0}(\omega)
+e^2{\cal F}_0(\chi) \big)
\big(S_{0}(-\omega)+e^2{\cal F}_0(\chi)\big)
\big].
\label{Omega}
\end{eqnarray}
Here ${\cal F}_0(\chi)$ is defined by Eq.~(\ref{cgfqpc}) with ${\cal T}_{|n\rangle}$ replaced
by the QPC transmission ${\cal T}$ in the regime when both quantum dots are empty, and
\begin{eqnarray}
Z_t(\omega)=\left[-i \omega \bar{C}
+e^2{\cal T}/\pi+1/Z_{\rm ext}(\omega)\right]^{-1} 
\label{Zt}
\end{eqnarray}
is the effective impedance seen by the QPC with 
$
\bar{C}
$ 
being the total effective capacitance seen by the QPC (see Fig.~\ref{fig:setup} (d)),
\begin{eqnarray}
\bar{C} =
\frac{C_L [ 3 C_0^2 + C_G (C_G+C_R) + 2 C_0 (2 C_G+C_R) ]}
{ 3 C_0^2 + (C_G+C_L) (C_G+C_R) + 2 C_0 \tilde C }.
\end{eqnarray} 
$S_\chi(\omega)$ is the non-symmetrized current noise spectral power given by the sum of the QPC shot noise and thermal
noise of the environment, $S_\chi(\omega)=S_{I, \chi}(\omega)+S_{\rm em}(\omega)$, 
\begin{eqnarray}
 S_{I , \chi}(\omega) 
\! &=& \! 
\frac{ e^2{\cal T} }{\pi}\left[ \frac{(\omega-eV_{\rm QPC})e^{-i\chi}}{1-e^{-\frac{\omega-eV_{\rm QPC}}{T}}}
+ \frac{(\omega+eV_{\rm QPC})e^{i\chi}}{1-e^{-\frac{\omega+eV_{\rm QPC}}{T}}} \right],
\nonumber \\
\label{SI1}
\\
S_{\rm em}(\omega)
\! &=& \!
{\rm Re}  \left[\frac{1}{Z_{\rm ext}(\omega)}\right]
\frac{2\omega}{1-e^{-\omega/T}}.
\label{Sem1}
\end{eqnarray}

\subsection{Generalized detailed balance and fluctuation theorem}

The tunneling rates (\ref{glr}-\ref{gld}) obey the generalized detailed balance
relations. In order to derive them, we first note that in equilibrium
the noise of the QPC environment satisfies the usual detailed balance relation 
\begin{eqnarray}
S_{\rm em}(\omega) {\rm e}^{-\beta \omega} = S_{\rm em}(-\omega),
\end{eqnarray}
where $\beta=1/T$ is the inverse temperature.
Since the QPC noise depends on the counting field $\chi$ it obeys a generalized
 detailed balance relation of the same type, 
\begin{eqnarray}
S_{I , -\chi+i \beta eV_{\rm QPC}}(\omega) {\rm e}^{-\beta \omega}
= 
S_{I , \chi}(-\omega). 
\label{eqn:re2}
\end{eqnarray}
With the aid of these two relations one can prove the identity for the
modified $P_{n,\chi}(E)$ functions:  
\begin{eqnarray}
P_{n, -\chi+i \beta eV_{\rm QPC}}(E)
 {\rm e}^{-\beta E}
=
P_{n, \chi}(-E) .
\label{edbp}
\end{eqnarray}
These identities combined with the properties of the Fermi function lead
to the generalized local detailed balance relations between the tunneling rates:
\begin{eqnarray}
\Gamma_{RL}(\chi)
\! &=& \!
\Gamma_{LR}(-\chi+i \beta eV_{\rm QPC})
\,
{\rm e}^{\beta (\Delta_L-\Delta_R)},
\label{eldb1}
\\
\Gamma_{DL}(\chi)
\! &=& \!
\Gamma_{LD}(-\chi+i \beta eV_{\rm QPC})
\,
{\rm e}^{\beta (\Delta_L+\mu_R)}, 
\label{eldb2}
\\
\Gamma_{DR}(\chi)
\! &=& \!
\Gamma_{RD}(-\chi+i \beta eV_{\rm QPC})
\,
{\rm e}^{\beta (\Delta_R+\mu_L)},
\label{eldb3} 
\end{eqnarray}
The relations (\ref{eldb1}-\ref{eldb3}) constitute
the first main result of our paper. 
They differ from the usual local detailed balance relations in one important point: 
in our model the tunneling rates depend on the counting field $\chi$ which has to be transformed as 
$\chi\to -\chi+ i \beta eV_{\rm QPC}$ to ensure the detailed balance symmetry. 

It is known~\cite{Lebowitz} that the local detailed balance is sufficient to prove the FT for an isolated DQD system (\ref{FT}). 
We will now demonstrate that the relations
(\ref{eldb1}-\ref{eldb3}) lead to the FT (\ref{fullft}) for the coupled system of
DQD and QPC. First, we note that the CGF of
the QPC  (\ref{cgfqpc}) has the property
\begin{eqnarray}
{\cal F}_n(\chi)={\cal F}_n(-\chi+i \beta eV_{\rm QPC}),
\label{ftqpc}
\end{eqnarray}
which is equivalent to the FT for an isolated QPC (\ref{FT}).
The same applies to the correction induced by the environment (\ref{cgfgauss}), 
\begin{eqnarray}
\delta {\cal F}_0(\chi)=\delta {\cal F}_0(-\chi+i \beta eV_{\rm QPC}).
\label{ftgauss}
\end{eqnarray}
Next, in the long-time limit  $\tau \gg 1/\Gamma_{mn}$
the characteristic function (\ref{eqn:cf}) approaches the asymptotic exponential behavior 
\begin{eqnarray}
{\cal Z}_\tau(\lambda,\chi)
\approx 
{\rm e}^{\tau {\cal F}(\lambda,\chi)}
\label{asympt}
\end{eqnarray}
where the CGF ${\cal F}$ is the eigenvalue of the matrix (\ref{M}) with the largest real part.
It is to be found from the equation
\begin{eqnarray}
\det
\left[
{\cal F} \mat{E}-\mat{M}(\lambda,\chi)
\right]=0.
\label{eigen2}
\end{eqnarray}
With the aid of the identities (\ref{eldb1}-\ref{ftgauss}) one can prove
the following property of the determinant  
\begin{eqnarray}
\det
\left[
{\cal F} \mat{E}-\mat{M}(-\lambda + i \beta eV_S,
-\chi + i \beta eV_{\rm QPC})
\right]
\nonumber \\
\! =\!
\det
\left[
{\cal F} \mat{E}-\mat{M}(\lambda,\chi)
\right]
\, . 
\label{eigen3}
\end{eqnarray}
Hence the function ${\cal F}(\lambda,\chi)$ obeys the symmetry  
\begin{eqnarray}
{\cal F}(\lambda, \chi)
=
{\cal F}(
- \lambda + i \beta eV_S
,
- \chi + i \beta eV_{\rm QPC}
). 
\label{eqn:FTCGF} 
\end{eqnarray}
Since in the long-time limit the characteristic function (\ref{asympt})
possesses the same symmetry, one can easily verify that the joint probability
distribution, given by the inverse Fourier transformation 
\begin{eqnarray}
P_\tau(q,q') = \int_{-\pi}^\pi \frac{d\lambda\, d\chi}{(2\pi)^2}\, e^{-i\lambda q-i\chi q'}\,{\cal Z}_\tau(\lambda,\chi),
\end{eqnarray}
satisfies the FT in the form (\ref{fullft}).
The proof of the FT is the second important result of our paper.

\subsection{Experimental verification of the FT for the coupled system}
 
Here we are going to discuss how one can verify Eq.~(\ref{fullft}) in the experiment.
The direct measurement of the joint probability distribution $P_\tau(q,q')$ 
is almost impossible with current experimental techniques 
because the tunneling of electrons through the QPC is a very fast process. 
Therefore, we propose a different route which requires only current and noise measurements.  
Namely, we propose to check the universal relations between nonlinear transport coefficients imposed by the FT~\cite{SaitoU,Atsumian,RSanchez}. 
The first one relates 
the lowest-order nonlinear conductance of the DQD with the correction to its noise
and reads
\begin{eqnarray}
T\frac{\partial^2I}{\partial V_S^2} 
\bigg|_{V_S=V_{\rm QPC}=0}
=  
\frac{\partial{\cal S}_{\rm DQD}}{\partial V_S}
\bigg|_{V_S=V_{\rm QPC}=0}.
\label{universal1}
\end{eqnarray}
Here $I$ is the current through the DQD, and ${\cal S}_{\rm DQD}$ is the DQD current noise. 
The identity (\ref{universal1}) can be derived from both the two terminal (\ref{FT}) and the four terminal (\ref{fullft}) versions of the FT~\cite{Nakamura}. 

One can derive a relation analogous to Eq. (\ref{universal1}) for the QPC current and noise. 
However, since the nonlinearity of the QPC $I-V$ curve is weak and easily smeared by noise and temperature, 
the latter relation is more difficult to verify in the experiment.

The second relation, which we propose to test, reads
\begin{eqnarray}
2T\frac{\partial^2 I_{\rm QPC}}{\partial V_S\partial V_{\rm QPC}} 
\bigg|_{V_S=V_{\rm QPC}=0}
&=&
\frac{\partial{\cal S}_{\rm QPC}}{\partial V_{S}}
\bigg|_{V_S=V_{\rm QPC}=0}
\nonumber \\ &&
+ \,
\frac{\partial{\cal S}_{\rm nl}}{\partial V_{\rm QPC}}
\bigg|_{V_S=V_{\rm QPC}=0}.
\label{universal2}
\end{eqnarray}
Here   
\begin{eqnarray}
{\cal S}_{\rm nl} = \int dt \big\langle \big( I(t) - \langle I\rangle \big)\big( I_{\rm QPC}(0) - \langle I_{\rm QPC}\rangle \big) \big\rangle
\end{eqnarray}
is the non-local cross correlation of the QPC and DQD noises.

Our model allows us to express all the derivatives, which appear in the Eqs. (\ref{universal1},\ref{universal2})
in terms of the tunneling rates $\Gamma_{mn}$ and the energy differences between the DQD
charge states $\Delta_m$. These parameters may be independently measured, which
provides additional check both for the FT and for our model.
The corresponding expressions read
\begin{eqnarray}
&& T\frac{\partial^2 I}{\partial V_S^2}
\! = \!
\frac{\partial{\cal S}_{\rm DQD}}{\partial V_S}
=
e^3\beta {\rm e}^{\beta (\Delta_L+\Delta_R)}
 \bigg\{
\Gamma_{LR}\Gamma_{LD}\Gamma_{RD}
\nonumber\\ &&\times\,
\frac{{\rm e}^{\beta (\Delta_L+\Delta_R)}(\Gamma_{LD} \Gamma_{RD}+\Gamma_{LR} \Gamma_{LD}-\Gamma_{LR} \Gamma_{RD})}{(A B)^2}
\nonumber\\ &&
-\,\Gamma_{LR}\Gamma_{LD}\Gamma_{RD}
\frac{{\rm e}^{2 \beta \Delta_R}\Gamma_{LD} \Gamma_{RD}}{(A B)^2}
\nonumber \\ &&
+\,2 T \Gamma_{LR}^2 
\frac{\Gamma_{LD}^2 \, \partial_{V_S} \Gamma_{RD} + \Gamma_{RD}^2 \,  \partial_{V_S} \Gamma_{LD}}{A B^2}
 \bigg\}, 
\label{gg}
\end{eqnarray}
where we have introduced the parameters
\begin{eqnarray}
A&=&{\rm e}^{\beta \Delta_R} + {\rm e}^{\beta \Delta_L} + {\rm e}^{\beta (\Delta_L+\Delta_R)},
\nonumber\\
B&=&{\rm e}^{\beta \Delta_R} \Gamma_{LD}\Gamma_{RD} + \Gamma_{LR}(\Gamma_{LD}+\Gamma_{RD}).
\end{eqnarray}
The nonlinear response of the QPC current reads, 
\begin{eqnarray}
&& 2 T \frac{\partial^2 I_{\rm QPC}}{\partial V_S\partial V_{\rm QPC}} = \frac{\partial{\cal S}_{\rm QPC}}{\partial V_S} 
\nonumber\\ &&
= e^3\frac{{\cal T}_{|L \rangle}-{\cal T}_{|D \rangle}}{\pi}
{\rm e}^{\beta (\Delta_L+2 \Delta_R)}
\nonumber\\ &&\times\,
\frac{\Gamma_{LR}(\Gamma_{LD}-\Gamma_{RD})+\Gamma_{LD}\Gamma_{RD}(2+ {\rm e}^{\beta \Delta_R})}{A^2B}
\nonumber\\ &&
+\, e^3\frac{{\cal T}_{|R \rangle}-{\cal T}_{|D \rangle}}{\pi}
{\rm e}^{\beta (2 \Delta_L+\Delta_R)}
\nonumber\\ &&\times\,
\frac{ \Gamma_{LR}(\Gamma_{LD}-\Gamma_{RD}) - \Gamma_{LD}\Gamma_{RD}{\rm e}^{\beta \Delta_R}(1+2{\rm e}^{-\beta \Delta_L})}{A^2B}.
\nonumber \\
\label{gc}
\end{eqnarray}
In Eqs. (\ref{gg},\ref{gc}) all tunnel rates are taken at zero QPC counting field, $\chi=0$ and zero bias voltages 
$V_{\rm QPC}=V_S=0$,
and can be directly measured in experiment.
Finally, we find that the cross correlation of the noises equals to zero,
\begin{eqnarray}
{\partial{\cal S}_{\rm nl}}/{\partial V_{\rm QPC}}\big|_{V_S=V_{\rm QPC}=0} = 0.
\end{eqnarray}
This result significantly simplifies the test of the relation (\ref{universal2}) making rather complicated measurement of the non-local noise unnecessary.

\section{Derivation of the master equation}
\label{sec:mastereq}

In this section we provide a detailed derivation of the master equation (\ref{master})
and the expressions for the tunneling rates (\ref{glr}-\ref{gld}). 
For simplicity, here we put the electron charge equal to 1, $e=1$,
and ignore electron spins. 

\subsection{Model Hamiltonian}
\label{sec:model}

We describe the system shown in Fig. \ref{fig:setup} (a-1) by the Hamiltonian
\begin{eqnarray}
H
&=&
H_{\rm res} + H_{\rm 12} + H_T \, , 
\label{Hfull}
\\
H_{\rm res}
&=& 
H_L + H_R  + H_1 + H_2  + H_{\rm em} + H_{\rm DQD} \,.
\end{eqnarray}
Here the Hamiltonians 
\begin{equation}
H_r 
=
\sum_{k}
\xi_{rk} \, {a_{r k}}^\dagger a_{r k}
\, , 
\;\;\;\;
(r=L,R)
\, ,
\end{equation}
describe the electrons in the left and right leads coupled to the DQD,
while the Hamiltonians 
\begin{eqnarray}
H_j=
\sum_{k}
\xi_{jk} \, 
{a_{j k}}^\dagger a_{j k},\;\;\; (j=1,2),
\end{eqnarray}
describe the leads $1$ and $2$ on both sides of the QPC.
The Hamiltonian of the double quantum dot $H_{\rm DQD}$ reads
\begin{eqnarray}
H_{\rm DQD} 
&=& 
\sum_{n=L,R,D} 
E_n |n \rangle \langle n|,
\label{HDQD}
\end{eqnarray}
where $E_n$ is the energy of the DQD charge state $|n\rangle$.

The electron transport through the QPC is described by the tunnel Hamiltonian 
\begin{eqnarray}
H_{\rm 12}
& = &
\, 
u
\, 
\sum_{k k'}
\left( t_{12} e^{i\theta} 
{a_{2 k}}^\dagger
a_{1 k'}
+
h.c.
\right),
\label{H12}
\end{eqnarray}
where $t_{12}$ is the hopping matrix element, 
$$u=\sum_{n=L,R,D} u_n | n \rangle \langle n|$$ 
is the operator
which accounts for the effect of the DQD charge state on the QPC tunneling, and  
$\theta$ is the phase induced by voltage fluctuations in the electromagnetic environment. 
If both quantum dots are unoccupied the QPC transmission probability reads 
$
{\cal T} = 4 \pi^2 t_{12}^2 \nu_1 \nu_2,
$
where $\nu_1,\nu_2$ are the densities of states in the leads 1 and 2.
If either left or right quantum dot is occupied the tunnel matrix element is suppressed and the transmission probability changes to
${\cal T}_{ | n \rangle }=u_n^2 \, {\cal T}$, $n=L,R,D$.  

The tunneling between the quantum dots and the left and right leads is described by the tunnel Hamiltonian 
\begin{eqnarray}
H_{\rm T}
&=&
\sum_{r=L,R}
\sum_k
\left( t_r\,e^{i \kappa_r \theta}{a_{rk}}^\dagger  
| \bar{r} \rangle \langle D| + h.c. \right)
\nonumber\\ &&
+
t_c \,e^{i \kappa_{dd} \theta} \, |R \rangle \langle L| + h.c. \, , 
\label{HT}
\end{eqnarray}
where $\bar{r}=L/R$ for $r=R/L$. 
Due to the capacitive coupling between QPC and DQD the phase $\theta$ also appears in this Hamiltonian. 
Finally, the electromagnetic environment with the impedance $Z_{\rm ext}(\omega)$
is modeled by a set of $LC$ resonators in the spirit of Caldera-Leggett model~\cite{Leggett}.
It is described by the Hamiltonian 
\begin{eqnarray}
H_{\rm em}
&=&
\frac{Q^2}{2 \bar{C}}
+
\sum_\alpha
\frac{Q_\alpha^2}{2 C_\alpha}
+
\frac{(\theta-\varphi_\alpha)^2-\theta^2}{2 L_\alpha}
\, ,
\end{eqnarray}
where the charge operator $Q$ is related to the phase operator $\theta$ by a commutation relation 
$[ \theta, Q ] = i$; $\varphi_\alpha$ are the coordinates, or phases, of the environment $LC$-oscillators 
and the corresponding charges are defined by the commutators
$[\varphi_\alpha, Q_\alpha]=i$. 
The spectral density of $LC$-resonators is chosen to reproduce
the impedance of the electric circuit $Z_{\rm ext}(\omega)$ 
\begin{eqnarray}
\sum_\alpha 
\frac{\omega_\alpha}{2 L_\alpha}
\left[
\delta (\omega-\omega_\alpha)
-
\delta (\omega+\omega_\alpha)
\right]
=\frac{\omega}{\pi}
{\rm Re}
\frac{1}{ Z_{\rm ext}(\omega)}. 
\end{eqnarray}

\subsection{Full counting statistics}
\label{sec:cf}

In this subsection we derive a formal expression for the CF (\ref{eqn:cf})
applying the FCS approach for the reduced density matrix~\cite{Braggio}.
We split the system into six reservoirs: five electronic ones, namely the leads 1,2,L,R and DQD, and 
the bosonic reservoir of $LC$-oscillators. The reservoirs are described by the Hamiltonians 
$H_1$, 
$H_2$, 
$H_L$, 
$H_R$, 
$H_{\rm DQD}$ and $H_{\rm em}$ respectively. 
At negative times $t<0$ all reservoirs are decoupled 
and the system is described by the factorized density matrix 
\begin{eqnarray}
\rho_0
&=&
\rho_{\rm L} \,
\rho_{\rm R} \,
\rho_{\rm 1} \,
\rho_{\rm 2} \,
\rho_{\rm em} \,
\sum_{n=L,R,D}
p_n(0)
| n \rangle
\langle n | 
\, ,
\\
\rho_r & = &
\frac{{\rm e}^{-(H_r-\mu_r N_r)/T}}{{\rm Tr}\,\left({\rm e}^{-(H_r-\mu_i N_r)/T}\right)},
\end{eqnarray}
where $N_r=\sum_{k} {a_{r k}}^\dagger a_{r k}$ ($r=L,R,1,2$)
are the  particle numbers in the $r$-th reservoir, $\mu_r$ are 
the chemical potentials with the chemical potential of the bosonic environments being equal to zero, 
$\mu_{\rm em}=0$, 
and $p_n$ are the initial occupation probabilities of the DQD states $|n\rangle$.   

\begin{figure}[ht]
\includegraphics[width=0.5 \columnwidth]{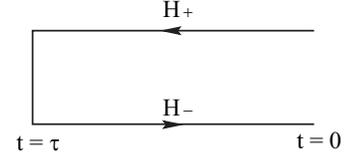}
\caption{
Keldysh contour $K$. 
The time runs from the right to the left. 
The total system evolves with the Hamiltonian 
$H_+$ for the forward branch (upper contour)
and 
$H_-$ for the backward branch (lower contour). 
}
\label{fig:2_0}
\end{figure}

At time $t=0$ the tunnel Hamiltonians $H_{12},H_{\rm T}$ are turned on and 
at $t=\tau$ they are turned off again. 
We introduce the initial $N_L^{(i)},N_1^{(i)}$ and final $N_L^{(f)},N_1^{(f)}$ numbers of the
electrons in the leads $L$ and $1$, and define the probability
$P_\tau(q,q')$ for $q$ electrons to tunnel out of the reservoir $L$
and $q'$ electrons -- out of the lead $1$ as follows
\begin{eqnarray}
P_\tau
(q,q')
&=&
\sum_{if}
|\langle f|{\rm e}^{-i H \tau}|i \rangle|^2 
\langle i| \rho_0 |i \rangle
\nonumber \\ 
&& \times
\delta_
{q, N_L^{(f)}-N_L^{(i)}}
\delta_
{q', N_1^{(f)}-N_1^{(i)}}
\, , 
\end{eqnarray}
where 
$| i \rangle$ and $| f \rangle$
are eigen states of the hamiltonian $H_{\rm res}$. 
The CF (\ref{eqn:cf}) is then expressed as, 
\begin{eqnarray}
{\cal Z}_\tau(\lambda,\chi)  
&=& \sum_{if} |\langle f|{\rm e}^{-i H \tau}|i \rangle|^2 \langle i|\rho_0| i\rangle 
\nonumber\\ &&\times\,
e^{ i \lambda \, \left( N_L^{(f)}-N_L^{(i)} \right) } 
e^{ i \chi \, \left( N_1^{(f)}-N_1^{(i)} \right) } 
\nonumber\\
&=& 
{\rm Tr} \! 
\left(
{\rm e}^{-i H_+ \tau}
\, \rho_0 \, 
{\rm e}^{i H_- \tau}
\right) \, .
\label{Z2}
\end{eqnarray}
Here we defined
\begin{eqnarray}
H_\pm &=& {\rm e}^{\mp i (\lambda N_L+ \chi N_1)/2} H {\rm e}^{\pm i (\lambda N_L+ \chi N_1)/2}.
\label{Hpm}
\end{eqnarray}
The only difference between these Hamiltonians and the original one (\ref{Hfull}) 
is the additional phase shift in the tunneling matrix amplitude. Namely, after the transformation (\ref{Hpm}) 
the tunnel Hamiltonians (\ref{H12},\ref{HT}) acquire the form
\begin{eqnarray}
H_{12 \, \pm}
& = & e^{i\theta}
t_{12} u 
\sum_{k k'}
\left(
{a_{2 k }}^\dagger
a_{1 k' }
{\rm e}^{\pm i \chi/2}
+
h.c.
\right), 
\end{eqnarray}
\begin{eqnarray}
H_{T \, \pm}
&=&
\sum_k
\left( t_L \, e^{\pm i \lambda} \, e^{i \kappa_{dd} \theta/2}
{a_{Lk}}^\dagger |R \rangle \langle D| + h.c. \right)
\nonumber\\ &&
+\, 
\sum_k
\left( t_R\,e^{-i \kappa_{dd} \theta/2}{a_{Rk}}^\dagger |L \rangle \langle D| + h.c. \right)
\nonumber\\ &&
+
\left( t_c \,e^{i \kappa_{dd} \theta} \, 
|R \rangle \langle L|
+ h.c. \right) .  
\label{HTpm}
\end{eqnarray}
Defining the Keldysh contour $K$, 
which starts at $t=0$,  propagates to $t=\tau$ and 
then returns back to $t=0$ (see Fig. \ref{fig:2_0}),
one can interpret $H_+$ as the system Hamiltonian on the forward branch of this contour,
while $H_-$ --- its Hamiltonian on the backward branch. 

Next, we split the Hamiltonians (\ref{Hpm}) into two parts $H_\pm=H_{0\pm}+H_{T\pm}$,
where the "non-perturbed" Hamiltonian consists of the reservoir part and the QPC hopping part, 
$$
H_{0 \pm}=H_{\rm res} + H_{{\rm 12} \pm} \, .
$$ 
Switching to the interaction representation we define
$
H_{T \pm} (t)_I
=
{\rm e}^{i H_{0 \pm} t}
H_{T \, \pm}
{\rm e}^{-i H_{0 \pm} t}
$
and rewrite CF in the form 
\begin{eqnarray}
{\cal Z}_\tau
=
{\rm Tr}
\left[
T_K \,
{\rm exp} \! 
\left (
-
i \int_K 
\!\!\!\!
dt \, 
H_{T \pm}(t)_I
\right )
{\rm e}^{iH_{0-} \tau}
{\rm e}^{-iH_{0+} \tau}
\rho_0 
\right ],
\nonumber \\
\end{eqnarray}
where $T_K$ is the time ordering operator along the Keldysh contour. 
By formally introducing the identity operator 
$\sum_{n=L,R,D} | n \rangle \langle n| =1$, 
expressed as the sum over the charge states of the DQD, into this formula  
we transform ${\cal Z}_\tau$ to the form Eq.~(\ref{eqn:cfp}) with the
occupation probabilities 
\begin{eqnarray}
p_n(\lambda,\chi;\tau)
&=&
{\rm Tr}
\biggl[
T_K \,
{\rm exp} \! 
\left(
-
i \int_K 
\!\!\!\!
dt \, 
H_{T \pm}(t)_I
\right )
\nonumber 
\\
& & \times
{\rm e}^{iH_{0-} \tau}
| n \rangle \langle n| 
{\rm e}^{-iH_{0+} \tau}
\rho_0 
\biggl]
\, , 
\label{p_}
\end{eqnarray}
which satisfy the normalization $\sum_n p_n(0,0;\tau)=1$. 

We find the probabilities (\ref{p}) performing the expansion in powers of the tunnel Hamiltonian $H_{T\pm}$ and using the real-time diagrammatic technique~\cite{Braggio,Schoeller,Koenig}. 
The details of this procedure are outlined in the next subsections.

\subsection{Zeroth order.}

If $H_{T\pm}=0$ the probabilities $p_n$ (\ref{p_}) read
\begin{eqnarray}
p_n^{(0)}(\tau)
=
\pi_n(\tau,0) 
\, 
p_n(0)
\, , 
\end{eqnarray}
where 
\begin{eqnarray}
\pi_n(\tau,0)
=
{\rm Tr}
\left(
{\rm e}^{i H_{n-} \tau}
{\rm e}^{-i H_{n+} \tau}
\, \rho_1 \, \rho_2 \, \rho_{\rm em}
\right) 
\end{eqnarray}
is the system propagator at fixed DQD charge $|n\rangle$. The Hamiltonians  
$
H_{n \pm}
=
H_1+H_2+H_{\rm em} 
+
\langle n|
H_{12 \pm}
| n \rangle
$
include the capacitive coupling between the DQD and the QPC. 
Employing the standard technique~\cite{Ambegaokar} 
we write the propagator $\pi^n_n(\tau,0)$ in the form
of the path integral over the fluctuating phase $\theta$ 
\begin{eqnarray}
\pi_n(\tau,0)
=
\int \! {\cal D}
[\theta]
\, 
\exp \left( i S_{\rm QPC}[\theta,u_n] + i S_{\rm em}[\theta] \right)
\, ,
\label{eqn:pathprop}
\end{eqnarray}
The action of the QPC is derived by tracing out electron degrees of freedom in leads 1 and 2 
and subsequent expansion of the result in powers of the tunnel matrix element $t_{12}$
(see Appendix \ref{sec:QPCaction}). The result of this procedure
is the nonlinear action, which makes the path integral (\ref{eqn:pathprop}) impossible to
evaluate exactly. To overcome this obstacle we assume that the impedance
of the electromagnetic environment is sufficiently small, $Z_S(\omega)\ll e^2/2\pi$,
at all relevant frequencies. In this regime the action $S_{\rm QPC}[\theta,u]$
becomes quadratic in phase fluctuations and takes the form, 
\begin{eqnarray}
S_{\rm QPC}[\theta,u]
& \approx &
\frac{i}{2} 
\int_K \!\! dt dt'
\, S_{I, \chi}(t,t') \, 
u(t) u(t') 
\nonumber \\ &&
\bigg(1
+ \theta(t) \theta(t')
-
\frac{\theta(t)^2+\theta(t')^2}{2}
\bigg). 
\label{AES}
\end{eqnarray}
It carries the information about the DQD charge state $|n\rangle$ through the
function $u(t)$, which, in general, depends on time and switches between the discreet values $u_n$. 
The function $S_{I,\chi}(t,t')$ is related to the QPC current noise and 
has the $2 \times 2$ matrix structure in Keldysh space   
\begin{eqnarray}
S_{I , \chi}(t,t')
=
\left(
\begin{array}{cc}
S_{I , \chi}^{++}(t,t') & S_{I , \chi}^{+-}(t,t') \\
S_{I , \chi}^{-+}(t,t') & S_{I , \chi}^{--}(t,t')
\end{array}
\right),
\end{eqnarray}
where the superscript $+$ indicates the forward branch of the Keldysh contour , while $-$ stands for the backward one.
In the limit of long measurement time 
$\tau \to \infty$ $S_{I , \chi}(t,t')$ depends on the time difference $t-t'$, which allows one
to make the Fourier transformation. Afterwards the greater and lesser components of this matrix take the form are expressed with Eq. (\ref{SI1}), 
\begin{eqnarray}
S_{I , \chi}^{-+}(\omega) = 
S_{I , \chi}^{+-}(-\omega) 
=
S_{I , \chi}(\omega). 
\label{eqn:re1}
\end{eqnarray}
Within our approximation the causal and anti-causal components do not depend on the counting field 
\begin{eqnarray}
S_{I , \chi}^{\pm\pm}(\omega)
\!&=&\!
\frac{
S_{I , 0}^{+-}(\omega)+S_{I , 0}^{-+}(\omega)
}{2}
\nonumber \\ &&
\pm
{\rm P}.
\int 
\frac{d \omega'}{2 \pi i}
\frac{
S_{I , 0}^{+-}(\omega')-S_{I , 0}^{-+}(\omega')
}{\omega-\omega'}, 
\label{cantic}
\end{eqnarray}
where the second part is the Cauchy principle value. 
The action of the electromagnetic environment reads 
\begin{eqnarray}
S_{\rm em}[\theta]
=
\int_K \!\! dt \frac{\bar{C}}{2} \dot{\theta}^2
+
\frac{i}{2}
\int_K \!\! dt dt' \theta(t) S_{\rm em}(t,t') \theta(t')
\, , 
\end{eqnarray}
where we have already traced out $Q$, $Q_\alpha$ and $\varphi_\alpha$. 
The lesser/greater component of the matrix $S_{\rm em}$ equals
$S_{\rm em}^{\pm \mp}(\omega)=S_{\rm em}(\mp \omega)$
[see Eq. (\ref{Sem1})], 
while causal/anti-causal components are given like Eq.~(\ref{cantic}).

Combining all these components, we express the total action $S_{\rm eff}=S_{\rm QPC}+S_{\rm em}$
in the form
\begin{eqnarray}
S_{\rm eff}[\theta,u]
=
S_{\rm QPC}[0,u]
+
\frac{1}{2}
\int_K  dt dt' \theta(t) D_\chi^{-1}(t-t') \theta(t'), 
\nonumber \\
\label{effectiveaction}
\end{eqnarray}
where
\begin{eqnarray}
D_\chi(\omega)^{-1}
&=&
\bar{C}
\left(
\begin{array}{cc}
-\omega^2 & 0 \\
0 & \omega^2
\end{array}
\right)
+
i
\left(
\begin{array}{cc}
 S_{\chi}^{++}(\omega) & -S_{\chi}^{+-}(\omega) \\
-S_{\chi}^{-+}(\omega) &  S_{\chi}^{--}(\omega)
\end{array}
\right)
\nonumber \\
&+&
i {\cal F}_0(\chi)
\left(
\begin{array}{cc}
1 & 0 \\
0 & 1
\end{array}
\right). 
\label{D-1} 
\end{eqnarray}
${\cal F}_0(\chi)$ is given by Eq. (\ref{cgfqpc}) with ${\cal T}_n$
replaced by ${\cal T}$, and
\begin{eqnarray}
S_{\chi}=S_{I,\chi}+S_{\rm em}
\, .  
\end{eqnarray}
Note that we have split the action (\ref{effectiveaction}) into the contribution, which depends only the DQD charging
state, i.e. on $u(t)$, and the contribution, which depends only the fluctuating phase $\theta$, but
its dependence on $u(t)$ is ignored  and we put $u\equiv 1$. This approximation
is justified as long the modulation of the QPC transmission probability by the DQD is small $|{\cal T}_n-{\cal T}|\ll{\cal T}$. 

4 components of the full photon matrix Green function $D_\chi$ are, 
\begin{eqnarray}
D_\chi^{\mp \pm}(\omega)
\! &=& \!
-i \, 
\frac{|Z_t(\omega)|^2}{\Omega_\chi(\omega)}
\frac{ S^{\mp \pm}_{\chi}(\omega) }{\omega^2}
\, . 
\\
D_\chi^{\pm \pm}(\omega)
\! &=& \!
-i \, 
\frac{|Z_t(\omega)|^2}{\Omega_\chi(\omega)}
\left(
\mp i \, \bar{C} + 
\frac{ 
S^{\mp \mp}_\chi(\omega) 
+
{\cal F}_0(\lambda)
}{\omega^2}
\right) ,
\nonumber \\
\label{eqn:gtheta}
\end{eqnarray}
where the function $\Omega_\chi(\omega)$ and the impedance $Z_t(\omega)$
are defined in Eqs. (\ref{Omega}) and (\ref{Zt}) respectively.

Let us now evaluate the first term of the effective action (\ref{effectiveaction}). 
In the limit $\tau\to\infty$ we get 
\begin{eqnarray}
i S_{\rm QPC}[0,u_n]
\! & \approx & \!
\tau
u_n^2 \, 
\frac{
S_{I , \chi}^{+-}(0)
+
S_{I , \chi}^{-+}(0)
-
S_{I , \chi}^{++}(0)
-
S_{I , \chi}^{--}(0)
}{2}
\nonumber \\
&=&
\tau {\cal F}_n(\chi)
\, ,
\end{eqnarray}
which is the CGF of the QPC~(\ref{cgfqpc})~\cite{Braggio2}. 
Hence the propagator (\ref{eqn:pathprop}) takes the form   
\begin{eqnarray}
\pi_n(\tau,0)
\approx
\int {\cal D}[\theta] {\rm e}^{i {S}_{\rm eff}}
=
{\rm e}^{\tau \, [{\cal F}_n(\chi) + \delta {\cal F}_0(\chi)]}, 
\end{eqnarray}
where $\delta {\cal F}_0(\chi) \propto -\ln\det[D^{-1}]$ is 
the correction to the QPC cumulant generating function,
which reads 
\begin{eqnarray}
\delta {\cal F}_0(\chi)
\approx
-\frac{1}{4 \pi} \int d \omega \ln \Omega_\chi(\omega)
\, . 
\label{cgfgauss} 
\end{eqnarray}
The correction  $\delta {\cal F}(\chi)$ is caused by weak the Coulomb blockade
effects in the QPC shunted by the impedance $Z_{\rm ext}(\omega)$ \cite{Kindermann}.  
By virtue of Eq. (\ref{eqn:re2}) and a similar identity 
$
S^{-+}_{\rm em} {\rm e}^{-\beta \omega}
=
S^{+-}_{\rm em}
$ 
one can show that
$
\Omega_{-\chi+i \beta V_{\rm QPC}}
=
\Omega_{\chi}.
$
Hence  
$\delta {\cal F}(\chi)$ satisfies the FT (\ref{ftgauss}).

\subsection{Second order: Photon assisted tunneling} 

\begin{figure}[ht]
\includegraphics[width=0.9 \columnwidth]{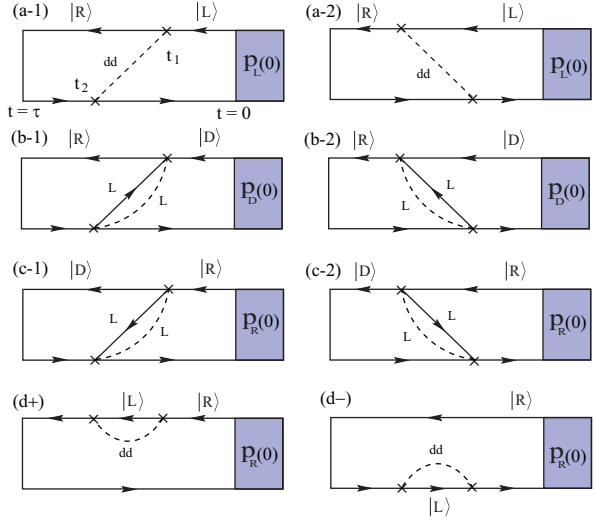}
\caption{
Typical diagrams contributing second order expansion in terms of the tunnel couplings. 
$\times$ represents hopping between the two dots $t_c$. 
Horizontal lines represent the dot state on the Keldysh upper and lower branches. 
Directed solid lines are reservoir lines. 
Dotted lines are correlation functions of the Gaussian phase fluctuations denoted by ${\rm e}^{i \kappa_{dd}^2 \Phi_\chi}$ etc. in the text. 
}
\label{fig:2}
\end{figure}

Let us  now discuss the second order expansion in $H_T$. 
Since the Gaussian effective action (\ref{effectiveaction}) is equivalent to some
effective bosonic bath, we adopt the corresponding diagrammatic rules proposed in Ref. ~\onlinecite{Koenig}. 
Figure \ref{fig:2} shows few typical diagrams. 
In these diagrams the time runs from the right to the left; the horizontal lines on the upper and lower branches of the Keldysh contour denote the states of the DQD; 
the directed solid lines, connecting the two branches, depict the Green functions of the reservoirs. 
The Fourier transformed lesser and greater components of the latter Green functions  read 
\begin{eqnarray}
\gamma_r^{\mp\pm}(\omega)
=
\Gamma_r
f(\mp \omega \pm \mu_r), \;\; (r=L,R).
\end{eqnarray}
where the rates $\Gamma_r = 2 \pi {t_r}^2 \nu_r$ determine the coupling between the dot $r$ and the lead $r$.

The sum of the two diagrams (a-1) and (a-2) reads
\begin{eqnarray}
P_{R}^{(a)}
=
\int^\tau_0 
\!\!\!\!
d t_2
d t_1
t_c^2  
{\pi}^{(a)}(\tau,t_2,t_1,0) \, {\rm e}^{i (E_R-E_L) (t_2-t_1)} p_L(0),
\nonumber \\
\label{eqn:a}
\end{eqnarray}
where the propagator is defined as follows
\begin{eqnarray}
\pi^{(a)}
&=&
{\rm Tr}
\biggl(
{\rm e}^{iH_{L-} t_1}
{\rm e}^{-i \kappa_{dd} \theta }
{\rm e}^{-iH_{R-} (t_1-\tau)}
{\rm e}^{-iH_{R+} (\tau-t_2)}
\nonumber \\
&& \times
{\rm e}^{i \kappa_{dd} \theta }
{\rm e}^{-iH_{L+} t_2}
\rho_1
\rho_2
\rho_{\rm em}
\biggl) 
\nonumber \\
\!& \approx &\!
\int \!\! {\cal D} [\theta] 
\exp
\left(
i
S_{\rm eff} \left[ \theta,u^{(a)} \right]
+
i
\int_K \!\!\! ds \, J^{(a)}(s) \, \theta(s)
\right) 
\nonumber \\
\!&=&\!
{\rm e}^
{i S_{\rm QPC} \left[ 0,u^{(a)} \right] + \tau \, \delta {\cal F}_0(\chi)}
\nonumber \\ && \times
\exp \left(
-
\frac{i}{2}
\int_K \!\! ds ds' \, 
J^{(a)}(s) \, D_\chi(s,s') \, J^{(a)}(s')
\right).
\nonumber \\
\label{path21}
\end{eqnarray}
The source  $J^{(a)}$ and the potential $u^{(a)}$ are expressed via the delta function $\delta$ and the step function $\theta$
defined on the Keldysh contour $K$: 
\begin{eqnarray}
J^{(a)}(s) \!&=&\! \kappa_{dd} \, (\delta(s,t_{2+})-\delta(s,t_{1-}) ) 
\, ,~
\\
u^{(a)}(s) \!&=&\! u_L + (u_R-u_L) \, (\theta(t_{1-},s)-\theta(s,t_{2+}))
.~~
\end{eqnarray}

Provided $\Gamma_r\ll T,eV$, we may use Markov approximation
and put
\begin{eqnarray}
i {\cal S}_{\rm QPC}
\left[0,u^{(a)}\right]
\approx 
{\cal F}_R(\chi) \, (\tau-t_2)
+
{\cal F}_L(\chi) \, t_1
\, . 
\end{eqnarray}
 The remaining double integral reduces to 
\begin{eqnarray}
&&
-
\frac{i}{2}
\int_K \!\! ds ds' \, 
J^{(a)}(s) \, D_\chi(s,s') \, J^{(a)}(s')
\nonumber \\
\! &=& \!
i
\kappa_{dd}^2 
\left[
D_\chi^{-+}(t_1,t_2)-
\frac{
D_\chi^{--}(t_1,t_1)+D_\chi^{++}(t_2,t_2)
}{2}
\right]
\nonumber \\
\! &=& \!
i\kappa_{dd}^2 \Phi_\chi(t_1-t_2), 
\label{eqn:PE}
\end{eqnarray}
where the function $\Phi_\chi(t)$ is defined in Eq. (\ref{Phi1}).
We note that at $\chi=0$ the function $\Phi_0(t)$ coincides
with the phase correlation function which appears in the so called
$P(E)$-theory (see e.g. Eq. (18) of Ref.~\onlinecite{Utsumi1}). 
Physically this function describes the back action of the QPC charge detector on the DQD through the photon assisted tunneling~\cite{Aguado} caused by the QPC shot noise. 

Thus after adopting Markov approximation the diagrams (a-1) and (a-2) give the result 
\begin{eqnarray}
\delta p_R^{(a)}(\tau) \approx 
\int_0^\tau \!\! dt
\, 
{\rm e}^{\tau \, \delta {\cal F}_0(\chi)
+
{\cal F}_R(\chi) \, (\tau-t)} \, 
\Gamma_{RL}(\chi) \, 
p_L^{(0)}(t)
\, ,
\end{eqnarray}
where $\Gamma_{RL}(\chi)$ is defined by Eq. (\ref{grl}).

Likewise the diagrams (b-1) and (b-2) give the contribution
\begin{eqnarray}
\delta p_R^{(b)}(\tau) 
\! &\approx& \!
\int_0^\tau \!\! dt
\, 
{\rm e}^{\tau \, \delta {\cal F}_0(\chi)
+{\cal F}_R(\chi) \, (\tau-t)} \, 
\Gamma_{RD}(\chi) 
{\rm e}^{-i \lambda}
\, 
\, 
p_D^{(0)}(t)
\, ,
\nonumber \\
\end{eqnarray}
where the factor ${\rm e}^{-i \lambda}$ 
containing the DQD counting field comes from the lines describing the electron tunneling through the DQD. 

The diagrams (c-1) and (c-2) describe the opposite transition processes and 
their contribution can be obtained from that of the diagrams (b-1) and (b-2) 
by interchanging the DQD states 
$D \leftrightarrow R$
and changing the direction of the reservoir line corresponding to 
$
\gamma^{-+}_L(\omega) \to 
\gamma^{+-}_L(-\omega)$.
In this way one arrives at the expression (\ref{gdr}) for the tunneling rate $\Gamma_{DR}$.

Let us now demonstrate that the function $P_{n,\chi}(E)$
satisfies the generalized detailed balance (\ref{edbp}).
It follows from the following property of the phase correlation
function $\Phi_\chi(t)$:
\begin{eqnarray}
&& 
\!\!\!\!
\!\!\!\!
\Phi_{-\chi+i \beta V_{\rm QPC}}(t)
\nonumber \\
\!&=&\!
i 
\int \!\! d \omega 
\frac{|Z_t(\omega)|^2}
{\omega^2 \Omega_{-\chi + i \beta V_{\rm QPC}}(\omega)}
\biggl(
{\cal F}_0(-\chi+i \beta V_{\rm QPC})
\nonumber \\ 
& &
+
S_{0}(\omega)
-
{\rm e}^{-i \omega t}
S_{-\chi + i \beta V_{\rm QPC}}(\omega)
\biggl)
\nonumber \\
\! &=& \!
i 
\int \!\! d \omega 
\frac{|Z_t(\omega)|^2}{\omega^2 \Omega_\chi(\omega)}
\nonumber \\ & & \times
\left(
S_{0}(\omega)
+
{\cal F}_0(\chi)
-
{\rm e}^{i \omega (t+i \beta)}
S_{\chi}(\omega)
\right)
\nonumber \\
&=&
\Phi_{\chi}(-t-i \beta)
\, ,
\end{eqnarray}
where we used, 
$
|Z_t(\omega)|=|Z_t(-\omega)|
$
and
$
\Omega_{\chi}(\omega)=\Omega_{\chi}(-\omega)
$
in addition to Eqs.~(\ref{eqn:re2}) and (\ref{eqn:re1}). 
The detailed balance relation (\ref{edbp}) can now be verified.

Let us now consider the diagrams (d+) and (d-), which define the
diagonal tunneling rate. 
The contribution of these
diagrams can be evaluated as follows 
\begin{eqnarray}
P_{R}^{(d)}(\tau)
\!&=&\!
-
t_c^2  
\left[
\int^\tau_0 \!\!\! d t_2 \int^{t_2}_0 \!\!\! d t_1 \, 
{\pi}^{(d+)}(\tau,t_2,t_1,0) 
\right.
\nonumber \\ &&
\left.
+
\int^\tau_0 \!\!\! d t_1 \int^{t_1}_0 \!\!\! d t_2 \, 
{\pi}^{(d-)}(\tau,t_2,t_1,0) 
\right]
\nonumber \\ &&
\times
{\rm e}^{i (E_L-E_R) (t_2-t_1)} p_R(0),
\label{eqn:c}
\end{eqnarray}
where for the propagator $\pi^{(d\pm)}$, are given by path integrals with the Gaussian action $S_{\rm eff}$ with the source fields
\begin{eqnarray}
J^{(d\pm)}(s) \!\!&=&\!\! \kappa_{dd} 
\, (\delta(s,t_{2\pm})-\delta(s,t_{1\pm}) ) \, , 
\\
u^{(d\pm)}(s) \!\!&=&\!\! u_R + (u_L-u_R) \, (\theta(t_{1\pm},s)-\theta(s,t_{2\pm})).~
\end{eqnarray}
Thus the sum of the diagrams (d+) and (d-) is evaluated to the expression 
\begin{eqnarray}
\delta p_R^{(d)}(\tau) 
\! &\approx& \!
-
\int_0^\tau \!\! dt
\, 
{\rm e}^{\tau \delta {\cal F}_0(\chi) + {\cal F}_R(\chi) \, (\tau-t)} \, 
{\Gamma}_{LR}(0) \, 
\, 
p_R^{(0)}(t),
\nonumber \\
\end{eqnarray}
where we neglected $\chi$ dependence of 
${\Gamma}_{LR}$ since it gives only small correction to 
${\cal F}_R(\chi)$. 

Other rates $\Gamma_{mn}$ are found analogously and also obey the generalized local detailed balance relations.
We note that at $\chi=0$ the master equation
(\ref{master}) reduces to the ordinary one, which conserves the probability $\sum_n p_n=1$.
Higher order expansions are systematically calculated in the same way. 
In this way, we obtain the integral form of the master equation (\ref{master}).

\section{Summary}
\label{sec:summary}

In conclusion, we have derived the full counting statistics for the coupled 
system of a double quantum dot (DQD) and a quantum point contact (QPC).
The non-trivial part of our analysis is the development of the
theoretical framework which accounts for both the 
quantum charge transport through the QPC and the classical
sequential tunneling through the DQD.
We find that the combined system should
be described by a generalized master equation, in which the tunneling rates
are modified by the QPC back action and depend on the QPC counting field.
We prove the generalized detailed balance relations between the tunneling
rates corresponding to time-reversed processes. These relations, in turn, lead
to the fluctuation theorem for the joint probability distribution of the
charges transfered through the DQD and the QPC. Finally, we derive
universal relations between the non-linear corrections to the currents and the noises,
which follow from the fluctuation theorem and can be verified in the 
experiment. 

We would like to emphasize that
in this paper we have limited ourselves by the regime of weak tunneling in the DQD. 
At stronger coupling between the quantum dots and the leads higher-order quantum corrections 
have to be taken into account and the master equation (\ref{master}) is not valid any more. In this case 
one has to deal with the FCS of non-equilibrium strongly correlated systems (see e.g. Ref. \onlinecite{UGS}), 
and we leave that problem for the future.

We thank 
Toshimasa Fujisawa, 
Hisao Hayakawa, 
Mattias Hettler, 
Kensuke Kobayashi, 
Bruno K\"ung
and 
Keiji Saito for helpful discussions.
This work has been supported by Strategic International Cooperative Program
of the Japan Science and Technology Agency (JST) and
by the German Science Foundation (DFG) and the Okasan-Katoh foundation.

\begin{appendix}

\section{Effective QPC action}
\label{sec:QPCaction}

The Keldysh action describing the QPC in Eq.~(\ref{eqn:pathprop}) is
defined as follows 
\begin{eqnarray}
{ S}_{\rm QPC}
\! &=& \!
\int_{K} \!\! dt 
\left \{
\sum_{j=1,2}
\sum_{k}
a_{j k}(t)^*
(i \partial_t - \xi_{jk})
a_{j k}(t)
\right. 
\nonumber \\
& &
-
t_{12} \, u(t) 
\sum_{k k'}
\left[
a_{2 k}(t)^*
a_{1 k'}(t) 
\, {\rm e}^{i \theta(t)+\chi(t)/2}
\right.
\nonumber \\
& &
\left. \left.
+ c.c. \right] \right \}, 
\label{eqn:actionqpc}
\end{eqnarray}
Here the QPC counting field $\chi(t)$ equals $\chi$ on the forward, and $-\chi$ on the backward branches of the Keldysh contour.
Integrating out the Grassmann fields $a_{j k}$ and taking care of the proper boundary conditions~\cite{Kamenev}, we arrive at the result 
\begin{eqnarray}
{ S}_{\rm QPC}
&=&
- i \, {\rm Tr} \, {\bf G}_{\chi_1}^{-1}
\, ,
\label{eqn:sqpc}
\\
{\bf G}_{\chi_1}^{-1}(t,t')
&=&
\left(
\begin{array}{cc}
g_1^{-1}(t,t') & -t_{12}(t,t')^*
\\
-t_{12}(t,t') & g_2^{-1}(t,t')
\end{array}
\right).
\label{G-1}
\end{eqnarray}
Here the times $t$ and $t'$ run over the whole Keldysh contour $K$, and the trace is understood 
as the time integral along $K$. 
The components of the matrix (\ref{G-1}) read 
\begin{eqnarray}
t_{12}(t,t')
&=&
t_{12} \, u(t) \, {\rm e}^{i \theta(t)+i \chi(t)/2}
\delta(t,t')
\, ,
\\
g_j(t,t') &=& \sum_k g_{j k}(t,t') 
\, ,
\\
g_{j k}^{-1}(t,t')
&=&
(i \partial_t -\xi_{jk}) \, \delta(t,t') 
\, .
\end{eqnarray}
Expanding the action~(\ref{eqn:sqpc}) up to second order in $t_{12}$, 
we obtain the Ambegaokar-Eckern-Sch\"on action~\cite{Ambegaokar} 
\begin{eqnarray}
{S}_{\rm QPC}
&=&
i
\int_K \!\! dt dt' \, 
u(t)u(t')
t_{12}^2
g_1(t,t')
g_2(t',t)
\nonumber \\
& & 
\times
{\rm e}^{
i[\theta(t)-\theta(t')]
+i[\chi(t)-\chi(t')]/2
}
\, .
\end{eqnarray}
Further expanding this action in $\theta$, we get the action (\ref{AES})
with 
\begin{eqnarray}
{S}_{I,\chi_1}(t,t')
\! &=& \!
u(t)u(t') \, 
t_{12}^2 \, 
g_1(t,t')
g_2(t',t)
\nonumber \\
& &
\times {\rm e}^{i[\chi(t)-\chi(t')]/2}
+ (t \leftrightarrow t')
\, .
\end{eqnarray}

\section{Nonlinear transport coefficients}
\label{sec:ntc}

In order to derive Eqs. (\ref{gg}) and (\ref{gc}) we solve the equation for the CGF (\ref{eigen2}) perturbatively in the counting fields $\chi,\lambda$ and bias voltages $V_S,V_{\rm QPC}$, and use the definitions
\begin{eqnarray}
&& I=-i\frac{\partial{\cal F}}{\partial\lambda},\;\; I_{\rm QPC}=-i\frac{\partial{\cal F}}{\partial\chi}, 
\nonumber\\
&& {\cal S}_{\rm DQD}=-\frac{\partial^2{\cal F}}{\partial\lambda^2},\;\; {\cal S}_{\rm QPC}=-\frac{\partial^2{\cal F}}{\partial\chi^2},\;\;
{\cal S}_{\rm nl}=-\frac{\partial^2{\cal F}}{\partial\lambda\partial\chi}. 
\nonumber
\end{eqnarray} 
Several additional symmetries make the calculations simpler. 
First, at $V_{\rm QPC}=0$, the local detailed balance for the DQD tunneling rates (Eqs.~(\ref{eldb1}-\ref{eldb3}) with $\chi=0$) are satisfied. 
Therefore at $V_S=V_{\rm QPC}=0$ the following relations hold
\begin{eqnarray}
\Gamma_{DL}(0) 
&=&
{\rm e}^{\beta \Delta_L} \Gamma_{LD}(0) ,
\nonumber\\
\frac{d \Gamma_{DL}(0)}{d V_S}
&=&
{\rm e}^{\beta \Delta_L} 
\left(
\frac{d \Gamma_{LD}(0)}{d V_S}
+
\beta
\frac{d \mu_R}{d V_S}
\Gamma_{LD}(0) 
\right),
\nonumber\\
\frac{d^2 \Gamma_{DL}(0)}{d V_S^2}
&=&
{\rm e}^{\beta \Delta_L} 
\left[
\frac{d^2 \Gamma_{LD}(0)}{d V_S^2}
+
2 \beta 
\frac{d \mu_R}{d V_S}
\frac{d \Gamma_{LD}(0)}{d V_S}
\right.
\nonumber \\ && 
\left. 
+
\left(
\beta
\frac{d \mu_R}{d V_S}
\right)^2
\Gamma_{LD}(0) 
\right].
\nonumber
\end{eqnarray}
Second, as long as we keep only the corrections $\propto\chi^2$, the dependence on $\chi$ is determined by the CGFs for the QPC (\ref{cgfqpc}),
which in this limit reads 
\begin{eqnarray}
{\cal F}_n \approx 
{\cal T}_{|n \rangle}/(2 \pi)
\left[
V_{\rm QPC} (i \chi)
+
T (i \chi)^2
+
\cdots
\right] \, .
\end{eqnarray}
The dependence of the tunneling rates on $\chi$ can be neglected, 
$
\Phi_\chi=\Phi_0
$
since at low bias, the QPC contribution is dominant.

Derivatives of the transition rates over $V_{\rm QPC}$ show how strongly the tunneling in the DQD is affected by the QPC shot noise. 
One can check that 
$
\left. 
\partial_{V_{\rm QPC}} S_{I,0}(\omega)
\right|_{V_{\rm QPC}=0}
=
0
$
and thus 
the first derivatives vanish, i.e. 
\begin{eqnarray}
\left. 
\partial_{V_{\rm QPC}} \Gamma_{mn}(0)
\right|_{V_{\rm QPC}=0}
=
0 \, . 
\end{eqnarray}
By utilizing these results, 
we arrive at Eqs. (\ref{gg}) and (\ref{gc})
after expanding the characteristic equation (\ref{eigen2})
in powers of $\chi$, $\lambda$, $V_S$ and $V_{\rm QPC}$ and 
comparing the coefficients
(A more systematic derivation for the higher cumulants in Ref.~\onlinecite{Flindt} is also applicable).

\end{appendix}

\end{document}